\documentclass[thmsa]{article}
\usepackage{sw20lart}

\input tcilatex
\QQQ{Language}{
British English
}

\begin{document}

\section{Introduction}

Spin has been shown to be amenable to a new treatment which provides new
insight into the origin of its matrix description[1-6]. This treatment is
based on the interpretation of quantum mechanics due to Land\'e [7-10]. The
ideas on which the treatment is based are general, and though the treatment
has initially been applied to spin systems, it should be of general
applicability. In this paper, we prove this by successfully applying this
approach to the description of polarization. We obtain new generalized
formulas for the polarization states and for the operators that enter into
the calculation of polarization expectation values.

This paper is organized as follows. In Section 2, we start off by giving the
basic theory underlying the new approach. In Section 3, we review the
connection between the differential and the matrix eigenvalue equation.
Section 4 is devoted to the application of the theory to the case of
polarization measurements. Thus, we derive the generalized probability
amplitudes for polarization measurements and calculate the probabilities
corresponding to them in Section 4.1. In Section 4.2 we give the general
formulas for the matrix operators of any observable whose eigenvalues depend
on the polarization direction. These formulas are straightaway used to
obtain the generalized polarization operator in Section 4.3. We consider
expectation values for polarization measurements in Section 4.4. In Section
4.5, we connect the present treatment with the standard approach by showing
how the generalized formulas we present here reduce to the standard ones in
the appropriate limit. We wrap up the paper in Section 5, with a Discussion
and then a Conclusion.

\section{Basic Theory}

The basic theory to be employed is a development of the Land\'e
interpretation of quantum mechanics[7-10]. According to this approach,
nature is inherently indeterministic, and quantum mechanics must necessarily
reflect this. Let a quantum system have the observables $A$, $B$ and $C$
which have the respective eigenvalue spectra $A_1$, $A_2$... $B_1$, $B_2$...
and $C_1$, $C_2...$ If the system is initially in a state corresponding to
the eigenvalue $A_i$, a measurement of $B$ yields any of the eigenvalues $%
B_j $ with probabilities determined by the probability amplitudes $\chi
(A_i,B_j).$ A measurement of $C$ results in one of the eigenvalues $C_j$
with probabilities determined by the probability amplitudes $\psi (A_i,C_j).$
If the system is initially in a state corresponding to the eigenvalue $B_i$,
a measurement of $C$ gives any of the eigenvalues $C_j$ with probabilities
determined by the probability amplitudes $\phi (B_i,C_j).$ The probability
amplitudes display a two-way symmetry contained in the Hermiticity condition

\begin{equation}
\psi (C_j,A_i)=\psi ^{*}(A_i,C_j).  \label{on1}
\end{equation}

These probability amplitudes are orthogonal:

\begin{equation}
\sum_j\psi ^{*}(A_i,C_j)\psi (A_k,C_j)=\delta _{ik}.  \label{fo4}
\end{equation}

As is to be expected of probability amplitudes pertaining to one system,
they are not independent and are connected to one another by the relation 
\begin{equation}
\psi (A_i,C_n)=\sum_j\chi (A_i,B_j)\phi (B_i,C_n).  \label{tw2}
\end{equation}
Each of the other two types of probability amplitudes can similarly be
expressed in terms of the other two sets. Relation (\ref{tw2}) is our
fundamental expression and most of what follows is based upon it.

\section{ Differential and Matrix Eigenvalue Equations}

In order to better understand the present approach, we review some results
regarding the connection between the differential and the matrix eigenvalue
equations for the same observable[1]. In the Land\'e formalism, the
eigenfunction is a probability amplitude connecting two states connected
together by a measurement. For this reason it is characterized by two
labels. One defines the state that obtains before measurement while the
other refers to the state that comes about as a result of the measurement.
The eigenvalue refers to the initial state, while the continuous variable in
terms of which the corresponding differential operator is framed describes
the final state. Thus, the eigenvalue equation for the operator $A(x)$ is
written out as

\begin{equation}
A(x)\psi (a_k,x)=a_k\psi (a_k,x).  \label{fi5}
\end{equation}
where $\psi (a_k,x)$ is an eigenfunction of $A$ with eigenvalue $a_k.$

We now transform this differential eigenvalue equation to a matrix
eigenvalue equation in the usual way. We use Eq. (\ref{tw2}) to express $%
\psi (a_k,x)\;$as an expansion:

\begin{equation}
\psi (a_k,x)=\dsum\limits_{j=1}^N\chi (a_k,B_j)\phi (B_j,x).  \label{si6}
\end{equation}
Going through the usual steps, we find the matrix eigenvalue equation

\begin{equation}
\left( 
\begin{array}{cccc}
A_{11}-a_k & A_{12} & ... & A_{1N} \\ 
A_{21} & A_{22}-a_k & ... & A_{2N} \\ 
... & ... & ... & ... \\ 
... & ... & ... & ... \\ 
A_{N1} & A_{N2} & ... & A_{NN}-a_k
\end{array}
\right) \left( 
\begin{array}{c}
\chi (a_k,B_1) \\ 
\chi (a_k,B_2) \\ 
... \\ 
... \\ 
\chi (a_k,B_N)
\end{array}
\right) =0,  \label{te10}
\end{equation}
where 
\begin{equation}
A_{mj}=\left\langle \phi (B_m,x)\left| A(x)\right| \phi (B_j,x)\right\rangle
.  \label{ni9}
\end{equation}
The expectation value of the arbitrary quantity $R(x)$ is given by

\begin{equation}
\left\langle R\right\rangle =\dsum\limits_{i=1}^N\dsum\limits_{j=1}^N\chi
^{*}(a_k,B_i)R_{ij}\chi (a_k,B_j)=[\chi (a_k)]^{\dagger }\left[ R\right]
[\chi (a_k)],  \label{el11}
\end{equation}
where

\begin{equation}
R_{ij}=\left\langle \phi (B_i,x)\left| R(x)\right| \phi (B_j,x)\right\rangle
,  \label{tw12}
\end{equation}

\begin{equation}
\left[ R\right] =\left( 
\begin{array}{cccc}
R_{11} & R_{12} & ... & R_{1N} \\ 
R_{21} & R_{22} & ... & R_{2N} \\ 
... & ... & ... & ... \\ 
... & ... & ... & ... \\ 
R_{N1} & R_{N2} & ... & R_{NN}
\end{array}
\right)  \label{fo14}
\end{equation}
and 
\begin{equation}
\lbrack \chi (a_k)]=\left( 
\begin{array}{c}
\chi (a_k,B_1) \\ 
\chi (a_k,B_2) \\ 
... \\ 
... \\ 
\chi (a_k,B_N)
\end{array}
\right) .  \label{th13}
\end{equation}
A very important point to note from Eq. (\ref{te10}) is that the elements of
the eigenvectors $[\chi (a_k)]$ of the operator $[A]$ are the probability
amplitudes $\chi (a_k,B_j).$ This is true even in the cases where the matrix
eigenvalue equation does not directly follow from a differential eigenvalue
equation. This is a very important result because it makes almost trivial
the task of obtaining the eigenvectors of the operator $[A].$ Basically, as
soon as the probability amplitudes for the measurement under consideration
are known, the eigenvectors are also known. We have extensively used this
result in our work on spin[1-5].

\section{Application to Polarization}

\subsection{Probability Amplitudes and Probabilities}

As we have done for spin, we shall use Eq. (\ref{tw2}) to obtain generalized
expressions for the probability amplitudes describing polarization
measurements. We shall then use these to obtain the matrix treatment of
polarization.

We consider a photon travelling in the $z$ direction. Its polarization is
measured with respect to the $x$ direction by means of the angle $\theta .$
If the photon is polarized at the angle $\theta $ with respect to the $x$
direction, then according to standard theory, its state is

\begin{equation}
\lbrack \chi (\theta ^{+})]=\left( 
\begin{array}{c}
\cos \theta \\ 
\sin \theta e^{i\alpha }
\end{array}
\right) ,  \label{fo14a}
\end{equation}
where $\alpha $ is the relative phase between the $x$ and $y$ components of
the electric field vector. If the polarization is at right angles to $\theta
,$ the state is

\begin{equation}
\lbrack \chi (\theta ^{-})]=\left( 
\begin{array}{c}
-\sin \theta \\ 
\cos \theta e^{i\alpha }
\end{array}
\right) .  \label{fi15}
\end{equation}

When $\theta =0$ and $\alpha =0,$ the state Eq. (\ref{fo14a}) gives $x$
polarization, while the state Eq. (\ref{fi15}) describes $y$ polarization.
On the other hand, if $\theta =\pi /4$ and $\alpha =\pi /2$, Eq. (\ref{fo14a}%
) describes right circular polarization while Eq. (\ref{fi15}) describes
left circular polarization.

In the spirit of the Land\'e approach, we should begin our treatment of
polarization by talking about probability amplitudes. Suppose that the
photon is initially polarized in the direction $\theta _a$ and its
polarization in the direction $\theta _b$ is then measured. We denote the
probability amplitude for this measurement by $\chi (\theta _a^{+},\theta
_b^{\pm }).$ Thus, the probability amplitude for finding the polarization to
be along the direction defined by $\theta _b$ is $\chi (\theta _a^{+},\theta
_b^{+})$, while that for finding it to be perpendicular to that direction is 
$\chi (\theta _a^{+},\theta _b^{-}).$ By the same token, if the polarization
is known to be initially perpendicular to the direction defined by $\theta
_a $, the probability amplitudes for finding it, upon measurement, to be
parallel and perpendicular to the direction defined by $\theta _b$ are $\chi
(\theta _a^{-},\theta _b^{+})$ and $\chi (\theta _a^{-},\theta _b^{-})$
respectively. The polarizations parallel and perpendicular to a given
direction define two orthogonal states of an observable which is
characterized by the angle defining that direction. We may expand the
probability amplitude $\chi $ using the fundamental relation Eq. (\ref{tw2}%
): thus, for example,

\begin{equation}
\chi (\theta _a^{+},\theta _b^{+})=\psi (\theta _a^{+},\theta _c^{+})\phi
(\theta _c^{+},\theta _b^{+})+\psi (\theta _a^{+},\theta _c^{-})\phi (\theta
_c^{-},\theta _b^{+})  \label{si16}
\end{equation}
where $\theta _c$ is an arbitrary third direction.

Owing to the fact that all the probability amplitudes occurring refer to
polarization-to-polarization measurements, we can use the symbol $\chi $ for 
$\psi $ and $\phi .$ Thus, we have

\begin{equation}
\chi (\theta _a^{+},\theta _b^{+})=\chi (\theta _a^{+},\theta _c^{+})\chi
(\theta _c^{+},\theta _b^{+})+\chi (\theta _a^{+},\theta _c^{-})\chi (\theta
_c^{-},\theta _b^{+}).  \label{se17}
\end{equation}

The other expansions are 
\begin{equation}
\chi (\theta _a^{+},\theta _b^{-})=\chi (\theta _a^{+},\theta _c^{+})\chi
(\theta _c^{+},\theta _b^{-})+\chi (\theta _a^{+},\theta _c^{-})\chi (\theta
_c^{-},\theta _b^{-}),  \label{ei18}
\end{equation}

\begin{equation}
\chi (\theta _a^{-},\theta _b^{+})=\chi (\theta _a^{-},\theta _c^{+})\chi
(\theta _c^{+},\theta _b^{+})+\chi (\theta _a^{-},\theta _c^{-})\chi (\theta
_c^{-},\theta _b^{+})  \label{ni19}
\end{equation}
and 
\begin{equation}
\chi (\theta _a^{-},\theta _b^{-})=\chi (\theta _a^{-},\theta _c^{+})\chi
(\theta _c^{+},\theta _b^{-})+\chi (\theta _a^{-},\theta _c^{-})\chi (\theta
_c^{-},\theta _b^{-}).  \label{tw20}
\end{equation}

We now need to obtain the explicit forms of these probability amplitudes. We
shall deduce these generalized probability amplitudes by considering Eqs. (%
\ref{th13}) and (\ref{tw2}).

According to the theory in Section 3, the elements of a matrix state are
probability amplitudes in their own right. This holds true for all
polarization state vectors. This means that from the standard state

\begin{equation}
\lbrack \chi (\theta _a^{+})]=\left( 
\begin{array}{c}
\cos \theta _a \\ 
\sin \theta _ae^{i\alpha _a}
\end{array}
\right)  \label{tw21}
\end{equation}
we can make the deductions

\begin{equation}
\chi (\theta _a^{+},\theta _f^{+})=\cos \theta _a  \label{tw22}
\end{equation}
and 
\begin{equation}
\chi (\theta _a^{+},\theta _f^{-})=\sin \theta _ae^{i\alpha _a},
\label{tw23}
\end{equation}
where $\theta _f$ is an unknown direction.

On the other hand, using 
\begin{equation}
\lbrack \chi (\theta _a^{-})]=\left( 
\begin{array}{c}
-\sin \theta _a \\ 
\cos \theta _ae^{i\alpha _a}
\end{array}
\right) ,  \label{tw24}
\end{equation}
we deduce that 
\begin{equation}
\chi (\theta _a^{-},\theta _f^{+})=-\sin \theta _a  \label{tw25}
\end{equation}
and 
\begin{equation}
\chi (\theta _a^{-},\theta _f^{-})=\cos \theta _ae^{i\alpha _a}.
\label{tw26}
\end{equation}

But suppose that the initial direction is $\theta _b$ instead of $\theta _a$%
. Then, from the standard states 
\begin{equation}
\lbrack \chi (\theta _b^{+})]=\left( 
\begin{array}{c}
\cos \theta _b \\ 
\sin \theta _be^{i\alpha _b}
\end{array}
\right)  \label{tw27}
\end{equation}
and 
\begin{equation}
\lbrack \chi (\theta _b^{-})]=\left( 
\begin{array}{c}
-\sin \theta _b \\ 
\cos \theta _be^{i\alpha _b}
\end{array}
\right) ,  \label{tw28}
\end{equation}
we deduce that 
\begin{equation}
\chi (\theta _b^{+},\theta _f^{+})=\cos \theta _b,  \label{tw29}
\end{equation}
\begin{equation}
\chi (\theta _b^{+},\theta _f^{-})=\sin \theta _be^{i\alpha _b},
\label{th30}
\end{equation}
\begin{equation}
\chi (\theta _b^{-},\theta _f^{+})=-\sin \theta _b  \label{th31}
\end{equation}
and 
\begin{equation}
\chi (\theta _b^{-},\theta _f^{-})=\cos \theta _be^{i\alpha _b}.
\label{th31a}
\end{equation}

Using Eq. (\ref{tw2}), together with Eq. (\ref{on1}), we can now eliminate
the unknown direction $\theta _f$ by expanding the probability amplitudes $%
\chi (\theta _a^{\pm },\theta _b^{\pm })$ and $\chi (\theta _a^{\mp },\theta
_b^{\pm })$ over the states corresponding to it. Thus, we obtain the
generalized probability amplitudes

\begin{equation}
\chi (\theta _a^{+},\theta _b^{+})=\cos \theta _a\cos \theta _b+\sin \theta
_a\sin \theta _be^{i(\alpha _a-\alpha _b)},  \label{th32}
\end{equation}

\begin{equation}
\chi (\theta _a^{+},\theta _b^{-})=-\cos \theta _a\sin \theta _b+\sin \theta
_a\cos \theta _be^{i(\alpha _a-\alpha _b)},  \label{th33}
\end{equation}
\begin{equation}
\chi (\theta _a^{-},\theta _b^{+})=-\sin \theta _a\cos \theta _b+\cos \theta
_a\sin \theta _be^{i(\alpha _a-\alpha _b)}  \label{th34}
\end{equation}
and 
\begin{equation}
\chi (\theta _a^{-},\theta _b^{-})=\sin \theta _a\sin \theta _b+\cos \theta
_a\cos \theta _be^{i(\alpha _a-\alpha _b)}.  \label{th35}
\end{equation}
In labelling the probability amplitudes, we have assumed that $\theta _a$
refers to the initial state. This is for two reasons. First, we used a
similar approach when we treated spin by this approach. On that occasion, we
found that the angle which explicitly appeared in the formulas for the
standard generalized results indeed referred to the initial state. Second,
the standard treatment defines only the initial state, and so the angle
which appears in the formulas must refer to that state.

We observe that by setting $\theta _b=\theta _f=0$, we obtain the standard
generalized results for the probability amplitudes. However, we now
recognise that these are not so generalized after all, and are only a
special case of the more generalized results Eqs. (\ref{th32}) - (\ref{th35}%
).

The probabilities corresponding to these probability amplitudes are 
\begin{eqnarray}
P(\theta _a^{+},\theta _b^{+}) &=&\left| \chi (\theta _a^{+},\theta
_b^{+})\right| ^2  \nonumber \\
&=&\cos ^2\theta _a\cos ^2\theta _b+\sin ^2\theta _a\sin ^2\theta _b+\frac
12\sin 2\theta _a\sin 2\theta _b\cos (\alpha _a-\alpha _b)  \label{th36}
\end{eqnarray}

\begin{equation}
P(\theta _a^{-},\theta _b^{-})=P(\theta _a^{+},\theta _b^{+})  \label{th36a}
\end{equation}
and 
\begin{eqnarray}
P(\theta _a^{+},\theta _b^{-}) &=&P(\theta _a^{-},\theta _b^{+})  \nonumber
\label{th38} \\
&=&\cos ^2\theta _a\sin ^2\theta _b+\sin ^2\theta _a\cos ^2\theta _b-\frac
12\sin 2\theta _a\sin 2\theta _b\cos (\alpha _a-\alpha _b).  \label{th37}
\end{eqnarray}

\subsection{Matrix Operators and States}

Consider the quantity $R(\theta _b)$, which is measured at the same time as
the polarization direction and whose values depend on the direction of
polarization. The value of $R(\theta _b)$ when the polarization is found to
be parallel to $\theta _b$ is denoted by $R_{+}$ while the value when the
polarization is found perpendicular to $\theta _b$ is called $R_{-}.$
Suppose further that the polarization is initially known to be along $\theta
_a$. Then the expectation value of $R(\theta _b)$ is

\begin{equation}
\left\langle R(\theta _b)\right\rangle =\left| \chi (\theta _a^{+},\theta
_b^{+})\right| ^2R_{+}+\left| \chi (\theta _a^{+},\theta _b^{-})\right|
^2R_{-}  \label{th39}
\end{equation}

We expand $\chi (\theta _a^{+},\theta _b^{+})$ and $\chi (\theta
_a^{+},\theta _b^{-})$ thus:

\begin{equation}
\chi ^{*}(\theta _a^{+},\theta _b^{\pm })=\chi ^{*}(\theta _a^{+},\theta
_c^{+})\chi ^{*}(\theta _c^{+},\theta _b^{\pm })+\chi ^{*}(\theta
_a^{+},\theta _c^{-})\chi ^{*}(\theta _c^{-},\theta _b^{\pm })  \label{fo40}
\end{equation}
and

\begin{equation}
\chi (\theta _a^{+},\theta _b^{\pm })=\chi (\theta _a^{+},\theta _c^{+})\chi
(\theta _c^{+},\theta _b^{\pm })+\chi (\theta _a^{+},\theta _c^{-})\chi
(\theta _c^{-},\theta _b^{\pm }).  \label{fo41}
\end{equation}

Hence,

\begin{eqnarray}
\left\langle R(\theta _b)\right\rangle &=&( 
\begin{array}{cc}
\chi _{11}^{*} & \chi _{12}^{*}
\end{array}
)\left( 
\begin{array}{cc}
R_{11} & R_{12} \\ 
R_{21} & R_{22}
\end{array}
\right) \left( 
\begin{array}{c}
\chi _{11} \\ 
\chi _{12}
\end{array}
\right)  \nonumber \\
\ &=&[\chi _{+}]^{\dagger }\left[ R\right] [\chi _{+}],  \label{fo42}
\end{eqnarray}
where 
\begin{equation}
\lbrack \chi _{+}]=\left( 
\begin{array}{c}
\chi _{11} \\ 
\chi _{12}
\end{array}
\right)  \label{fo43}
\end{equation}
and 
\begin{equation}
\lbrack R]=\left( 
\begin{array}{cc}
R_{11} & R_{12} \\ 
R_{21} & R_{22}
\end{array}
\right) .  \label{fo44}
\end{equation}
Here,

\begin{equation}
\chi _{11}=\chi (\theta _a^{+},\theta _c^{+})\ \;\;\text{and}\;\;\;\chi
_{12}=\chi (\theta _a^{+},\theta _c^{-}),\;\;  \label{fo45}
\end{equation}
while

\begin{equation}
R_{11}=\chi ^{*}(\theta _c^{+},\theta _b^{+})\chi (\theta _c^{+},\theta
_b^{+})R_{+}+\chi ^{*}(\theta _c^{+},\theta _b^{-})\chi (\theta
_c^{+},\theta _b^{-})R_{-},  \label{fo46}
\end{equation}

\begin{equation}
R_{12}=\chi ^{*}(\theta _c^{+},\theta _b^{+})\chi (\theta _c^{-},\theta
_b^{+})R_{+}+\chi ^{*}(\theta _c^{+},\theta _b^{-})\chi (\theta
_c^{-},\theta _b^{-})R_{-},  \label{fo47}
\end{equation}
\begin{equation}
R_{21}=\chi ^{*}(\theta _c^{-},\theta _b^{+})\chi (\theta _c^{+},\theta
_b^{+})R_{+}+\chi ^{*}(\theta _c^{-},\theta _b^{-})\chi (\theta
_c^{+},\theta _b^{-})R_{-}  \label{fo48}
\end{equation}
and

\begin{equation}
R_{22}=\chi ^{*}(\theta _c^{-},\theta _b^{+})\chi (\theta _c^{-},\theta
_b^{+})R_{+}+\chi ^{*}(\theta _c^{-},\theta _b^{-})\chi (\theta
_c^{-},\theta _b^{-})R_{-}.  \label{fo49}
\end{equation}

If on the other hand the initial state is such that the polarization is at
right angles to $\theta _a$, the state that appears in Eq. (\ref{fo42})
becomes 
\begin{equation}
\lbrack \chi _{-}]=\left( 
\begin{array}{c}
\chi _{21} \\ 
\chi _{22}
\end{array}
\right) =\left( 
\begin{array}{c}
\chi (\theta _a^{-},\theta _c^{+}) \\ 
\chi (\theta _a^{-},\theta _c^{-})
\end{array}
\right) .  \label{fi50}
\end{equation}

Using the generalized probability amplitudes Eqs. (\ref{th32}) - (\ref{th35}%
), we find that the generalized polarization states, denoted by $[\chi
_{\theta _a^{\pm }}],$are 
\begin{equation}
\lbrack \chi _{\theta _a^{+}}]=\left( 
\begin{array}{c}
\cos \theta _a\cos \theta _c+\sin \theta _a\sin \theta _ce^{i(\alpha
_a-\alpha _c)} \\ 
-\cos \theta _a\sin \theta _c+\sin \theta _a\cos \theta _ce^{i(\alpha
_a-\alpha _c)}
\end{array}
\right)  \label{fi51}
\end{equation}
and

\begin{equation}
\lbrack \chi _{\theta _a^{-}}]=\left( 
\begin{array}{c}
-\sin \theta _a\cos \theta _c+\cos \theta _a\sin \theta _ce^{i(\alpha
_a-\alpha _c)} \\ 
\sin \theta _a\sin \theta _c+\cos \theta _a\cos \theta _ce^{i(\alpha
_a-\alpha _c)}
\end{array}
\right) .  \label{fi52}
\end{equation}

In order to obtain Eqs. (\ref{fi51}) and (\ref{fi52}) we have replaced $b$
by $c$ in Eqs. (\ref{th32}) - (\ref{th35}). The states are orthogonal and
are each normalized to unity.

Using the expressions Eqs. (\ref{th32}) - (\ref{th35}) for the generalized
probability amplitudes with the arguments changed appropriately, so that $%
\theta _a\rightarrow \theta _c$ and $\theta _b$ remains unchanged, we obtain

\begin{eqnarray}
R_{11} &=&[\cos ^2\theta _c\cos ^2\theta _b+\sin ^2\theta _c\sin ^2\theta
_b+\frac 12\sin 2\theta _c\sin 2\theta _b\cos (\alpha _c-\alpha _b)]R_{+} 
\nonumber \\
&&+[\cos ^2\theta _c\sin ^2\theta _b+\sin ^2\theta _c\cos ^2\theta _b-\frac
12\sin 2\theta _c\sin 2\theta _b\cos (\alpha _c-\alpha _b)]R_{-},
\label{fi53}
\end{eqnarray}

\begin{eqnarray}
R_{12} &=&[-\frac 12\sin 2\theta _c\cos 2\theta _b+\frac 12\sin 2\theta
_b\cos 2\theta _c\cos (\alpha _c-\alpha _b)  \nonumber \\
&&+\frac i2\sin 2\theta _b\sin (\alpha _c-\alpha _b)]R_{+}+[\frac 12\sin
2\theta _c\cos 2\theta _b  \nonumber \\
&&-\frac 12\sin 2\theta _b\cos 2\theta _c\cos (\alpha _c-\alpha _b)-\frac
i2\sin 2\theta _b\sin (\alpha _c-\alpha _b)]R_{-},  \nonumber \\
&&  \label{fi54}
\end{eqnarray}

\begin{eqnarray}
R_{21} &=&[-\frac 12\sin 2\theta _c\cos 2\theta _b+\frac 12\sin 2\theta
_b\cos 2\theta _c\cos (\alpha _c-\alpha _b)  \nonumber \\
&&+\frac i2\sin 2\theta _b\sin (\alpha _c-\alpha _b)]R_{+}+[\frac 12\sin
2\theta _c\cos 2\theta _b  \nonumber \\
&&\ \ -\frac 12\sin 2\theta _b\cos 2\theta _c\cos (\alpha _c-\alpha
_b)-\frac i2\sin 2\theta _b\sin (\alpha _c-\alpha _b)]R_{-}  \nonumber \\
&&  \label{fi55}
\end{eqnarray}
and 
\begin{eqnarray}
R_{22} &=&[\sin ^2\theta _c\cos ^2\theta _b+\cos ^2\theta _c\sin ^2\theta
_b-\frac 12\sin 2\theta _c\sin 2\theta _b\cos (\alpha _c-\alpha _b)]R_{+} 
\nonumber \\
&&+[\sin ^2\theta _c\sin ^2\theta _b+\cos ^2\theta _c\cos ^2\theta _b+\frac
12\sin 2\theta _c\sin 2\theta _b\cos (\alpha _c-\alpha _b)]R_{-}.  \nonumber
\\
&&  \label{fi56}
\end{eqnarray}

\subsection{Polarization Operator and its Eigenvectors}

Suppose that the quantity $R$ is the polarization itself, which we denote by 
$p$. We agree that if measurement finds the polarization parallel to $\theta
_b$, we assign the value $+1$ to this observable. If the polarization is
found perpendicular to $\theta _b$, we assign the value $-1$. Then the
operator Eq. (\ref{fo44}) is found have the elements 
\begin{equation}
p_{11}=\cos 2\theta _c\cos 2\theta _b+\sin 2\theta _c\sin 2\theta _b\cos
(\alpha _c-\alpha _b),  \label{fi57}
\end{equation}

\begin{equation}
p_{12}=-\sin \theta _c\cos \theta _b+\cos 2\theta _c\sin 2\theta _b\cos
(\alpha _c-\alpha _b)+i\sin 2\theta _b\sin (\alpha _c-\alpha _b),
\label{fi58}
\end{equation}

\begin{equation}
p_{21}=-\sin \theta _c\cos \theta _b+\cos 2\theta _c\sin 2\theta _b\cos
(\alpha _c-\alpha _b)-i\sin 2\theta _b\sin (\alpha _c-\alpha _b)
\label{fi59}
\end{equation}
and 
\begin{equation}
p_{22}=-\cos 2\theta _c\cos 2\theta _b-\sin 2\theta _c\sin 2\theta _b\cos
(\alpha _c-\alpha _b).  \label{si60}
\end{equation}

According to the reasoning in Section 3, the eigenvectors of this operator
are

\begin{equation}
\lbrack \xi _{+}]=\left( 
\begin{array}{c}
\chi (\theta _b^{+},\theta _c^{+}) \\ 
\chi (\theta _b^{+},\theta _c^{-})
\end{array}
\right) =\left( 
\begin{array}{c}
\cos \theta _b\cos \theta _c+\sin \theta _b\sin \theta _ce^{i(\alpha
_b-\alpha _c)} \\ 
-\cos \theta _b\sin \theta _c+\sin \theta _b\cos \theta _ce^{i(\alpha
_b-\alpha _c)}
\end{array}
\right)  \label{si61}
\end{equation}
for eigenvalue $\lambda =1,$ and 
\begin{equation}
\lbrack \xi _{-}]=\left( 
\begin{array}{c}
\chi (\theta _b^{-},\theta _c^{+}) \\ 
\chi (\theta _b^{-},\theta _c^{-})
\end{array}
\right) =\left( 
\begin{array}{c}
-\sin \theta _b\cos \theta _c+\cos \theta _b\sin \theta _ce^{i(\alpha
_b-\alpha _c)} \\ 
\sin \theta _b\sin \theta _c+\cos \theta _b\cos \theta _ce^{i(\alpha
_b-\alpha _c)}
\end{array}
\right)  \label{si62}
\end{equation}
for $\lambda =-1.$ Direct calculation shows that indeed, the eigenvalue
equations 
\begin{equation}
\lbrack p][\xi _{\pm }]=\pm [\xi _{\pm }]  \label{si63}
\end{equation}
are satisfied.

\subsection{Expectation Values}

The expectation value of the polarization operator itself is now easily
obtained. Suppose that the polarization state that obtains before
measurement corresponds to polarization in the direction $\theta _a$, and
the polarization is measured in the direction $\theta _b.$ Then this
expectation value is

\begin{eqnarray}
\left\langle p\right\rangle _{+} &=&P(\theta _a^{+},\theta
_b^{+})(+1)+P(\theta _a^{+},\theta _b^{-})(-1)  \nonumber \\
&=&\cos 2\theta _a\cos 2\theta _b+\sin 2\theta _a\sin 2\theta _b\cos (\alpha
_a-\alpha _b)  \label{si64}
\end{eqnarray}

Similarly, 
\begin{eqnarray}
\left\langle p\right\rangle _{-} &=&P(\theta _a^{-},\theta
_b^{+})(+1)+P(\theta _a^{-},\theta _b^{-})(-1)  \nonumber \\
\ &=&-\cos 2\theta _a\cos 2\theta _b-\sin 2\theta _a\sin 2\theta _b\cos
(\alpha _a-\alpha _b)  \label{si65}
\end{eqnarray}
for the cases when the polarization is initially parallel and perpendicular
to the angle $\theta _a$ respectively.

\subsection{Connection With Standard Formulas}

The standard results are obtained from the current generalized results by
setting the final angle equal to zero. Thus if we set $\theta _b=0,$ $\alpha
_b=0$ in Eqs. (\ref{th32}) - (\ref{th33}), we obtain the standard formulas 
\begin{equation}
\chi _a^{+}=\cos \theta _a  \label{si66}
\end{equation}

\begin{equation}
\chi _a^{-}=\sin \theta _ae^{i\alpha _a},  \label{si67}
\end{equation}
The other two formulas which result from Eqs. (\ref{th34}) - (\ref{th35})
can be obtained from Eqs. (\ref{si66}) - (\ref{si67}) by having $\theta
_a\rightarrow \theta _a+\pi /2$, because this is the condition that defines
a state orthogonal to the one corresponding to $\theta _a$. Thus, we get 
\begin{equation}
\chi _{a\perp }^{+}=-\sin \theta _a  \label{si68}
\end{equation}
and 
\begin{equation}
\chi _{a\perp }^{-}=\cos \theta _ae^{i\alpha _a},  \label{si69}
\end{equation}
where we have used the notation $a\perp $ to indicate the quantities
corresponding to the direction perpendicular to $\theta _a.$

The polarization states in this limit are 
\begin{equation}
\lbrack \chi _{\theta _a^{+}}]=\left( 
\begin{array}{c}
\cos \theta _a \\ 
\sin \theta _ae^{i\alpha _a}
\end{array}
\right)  \label{se70}
\end{equation}
and

\begin{equation}
\lbrack \chi _{\theta _a^{-}}]=\left( 
\begin{array}{c}
-\sin \theta _a \\ 
\cos \theta _ae^{i\alpha _a}
\end{array}
\right) .  \label{se71}
\end{equation}

In order to obtain the standard polarization operator and its eigenvectors,
we have to set $\theta _c=0$, $\alpha _c=0$ in Eqs. (\ref{fi57}) - (\ref
{si62}) because this is the angle that now corresponds to the final
direction. The polarization operator then becomes

\begin{equation}
\lbrack p]=\left( 
\begin{array}{cc}
\cos 2\theta _b & \sin \theta _be^{i(\alpha _a-\alpha _b)} \\ 
\sin \theta _be^{-i(\alpha _a-\alpha _b)} & -\cos 2\theta _b
\end{array}
\right)  \label{se72}
\end{equation}
while its eigenvectors become 
\begin{equation}
\lbrack \xi _{+}]=\left( 
\begin{array}{c}
\cos \theta _b \\ 
\sin \theta _be^{i(\alpha _b-\alpha _c)}
\end{array}
\right)  \label{se73}
\end{equation}
for eigenvalue $+1$ and 
\begin{equation}
\lbrack \xi _{-}]=\left( 
\begin{array}{c}
-\sin \theta _b \\ 
\cos \theta _be^{i(\alpha _b-\alpha _c)}
\end{array}
\right)  \label{se74}
\end{equation}
for eigenvalue $-1.$

\section{Conclusion}

In this paper, we have presented generalized formulas for the treatment of
polarization. These should prove useful in the description of phenomena in
which polarization plays a part by giving us increased freedom of
description. Particularly noteworthy in this regard are the formulas for the
elements of the matrix form of the operator of any observable whose
eigenvalues depend on the polarization.

The results in this paper demonstrate once again the utility of the approach
to quantum mechanics due to Land\'e. The debate on the interpretation of
quantum mechanics continues unabated, and sometimes ascends to the
metaphysical. At other times it seems to be redundant, as when variants of
the theory are produced which yield only the standard results while allowing
a different interpretation of the foundations. The present approach at least
produces generalizations and new results which other approaches appear
unable to do. To this extent, it justifies itself. We have no doubt that
these generalizations will in future lead to important advances in both the
understanding and the application of quantum mechanics.

\section{References}

1. Mweene H. V., ''Derivation of Spin Vectors and Operators From First
Principles'', quant-ph/9905012

2. Mweene H. V., ''Generalized Spin-1/2 Operators and Their Eigenvectors'',
quant-ph/9906002

3. Mweene H. V., ''Vectors and Operators for Spin 1 Derived From First
Principles'', quant-ph/9906043

4. Mweene H. V., ''Alternative Forms of Generalized Vectors and Operators
for Spin 1/2'', quant-ph/9907031

5. Mweene H. V., ''Spin Description and Calculations in the Land\'e
Interpretation of Quantum Mechanics'', quant-ph/9907033

6. Mweene H. V., ''A New Approach to the Treatment of Systems of Compounded
Angular Momentum'', quant-ph/9907082

7. Land\'e A., ''From Dualism To Unity in Quantum Physics'', Cambridge
University Press, 1960.

8. Land\'e A., ''New Foundations of Quantum Mechanics'', Cambridge
University Press, 1965.

9. Land\'e A., ''Foundations of Quantum Theory,'' Yale University Press,
1955.

10. Land\'e A., ''Quantum Mechanics in a New Key,'' Exposition Press, 1973.

\end{document}